\documentclass[aps,pra,showpacs,twocolumn,floatfix]{revtex4}
\usepackage{graphicx} % Include figure files
\usepackage{amsmath}
\usepackage{bm}
\usepackage{dcolumn}% Align table columns on decimal point
\begin{document}

\title{ Hyperfine structure
of the metastable $^3P_2$ state of alkaline earth atoms as an
accurate probe of nuclear magnetic octupole moments}

\author{  K. Beloy and A. Derevianko }
 \affiliation{Physics Department, University of
Nevada, Reno, Nevada  89557}
\author{ W. R. Johnson }
 \affiliation{ Physics Department, University of Notre Dame, Notre
 Dame, Indiana 46566 }

\date{\today}
\begin{abstract}
Measuring the hyperfine structure (HFS) of long-lived $^3P_2$
states of divalent atoms may offer the opportunity of extracting
relatively unexplored nuclear magnetic octupole and electric
hexadecapole moments. Here, using relativistic many-body methods
of atomic structure and the nuclear shell model, we evaluate the
effect of these higher nuclear moments on the hyperfine structure.
We find that the sensitivity of HFS interval measurements in
$^{87}$Sr needed to reveal the perturbation caused by the nuclear
octupole moment is on the order of kHz.  Results of similar
analyses for $^{9 }$Be, $^{25}$Mg, and $^{43}$Ca are also
reported.
\end{abstract}

\pacs{06.30.Ft, 32.10.Dk, 31.25.-v} \maketitle

%\email{andrei@unr.edu}

% 2006 PACS
%32.10.Dk Electric and magnetic moments, polarizability
%06.30.Ft Time and frequency
%31.25.-v Electron correlation calculations for atoms and molecules

\section{Introduction}
Progress in cooling and trapping techniques has enabled a number
of advances in precision measurements of atomic properties.
Compared to more traditional beam spectroscopy, trapping atoms or
ions increases interrogation times and enhances spectral
resolution. This is a well-recognized experimental trend spanning
from the searches for permanent electric-dipole moments to atomic
clocks. Here, we point out that measuring the hyperfine structure
of long-lived $^3P_2$ states of cold divalent atoms may facilitate
extracting the so far relatively unexplored nuclear magnetic
octupole moments.

A nucleus may be described as a collection of electromagnetic
moments. Generally, a nucleus of spin $I$ possesses $2I$
moments beyond the monopole. Most well-known are the magnetic dipole
($\mu$) and electric quadrupole ($Q$) moments. These two leading
moments have been studied extensively and have contributed to the
understanding of nuclear forces.   For example, discovery of the
quadrupole moment of the deuteron led to introducing nuclear
tensor forces~\cite{Ram69}. Here, we are interested in the next two
moments in the multipolar hierarchy: nuclear magnetic
octupole ($\Omega$) and electric hexadecapole ($\Pi$) moments.

Electromagnetic fields of various nuclear moments  perturb the
motion of atomic electrons and reveal themselves in the hyperfine
structure (HFS) of atomic and molecular levels. The resulting
hyperfine intervals are conventionally parameterized in terms of
the HFS constants $A, B, C, D,\cdots$, each being proportional to
the relevant nuclear multipolar moment. The perturbation becomes
weaker as the rank of the multipole moment increases, and
detecting the influence of higher multipole moments on the HFS
intervals requires increasingly longer
interrogation times. Also, to observe the effects of higher rank
multipoles on the HFS structure one is forced to work with
electronic states of relatively high angular momenta $J$ since the
$2^k$-pole moment only contributes in first order to the HFS
structure of levels with $J \ge k/2$.  These two factors have
limited the determination of the octupole HFS constants to a small
number of isotopes of Cl~\cite{Sch57}, Ga~\cite{DalHol54},
Br~\cite{BroKin66}, In~\cite{EckKus57}, V~\cite{ChiPouGoo79V},
Eu~\cite{Chi91}, Lu~\cite{BreButRup85}, and Hf~\cite{JinWakIna95}.
All of the above measurements have been carried out either on the
ground state or on metastable states.
 The most recent octupole moment measurement was carried on the short-lived
$6p_{3/2}$ state of $^{133}$Cs~\cite{GerDerTan03}. Except for the case of the monovalent
Cs atom, deducing octupole moments from  measured HFS constants
presents a formidable challenge, owing to the fact that the prerequisite
atomic-structure calculations become inaccurate for multi-valent atoms.
As a result, previous analyses focused primarily on {\em ratios} of octupole moments
for various isotopes of the same atom, because the atomic-structure factor
cancels out when ratios of HFS constants are formed. By contrast,
the divalent alkaline-earth atoms considered here potentially yield direct values
of the nuclear octupole moments.

\begin{table*}[ht]
\caption{\label{Tab:Isotopes} Nuclear parameters and lifetimes of
the $^3P_2$ states of the stable isotopes of Be, Mg, Ca, and Sr. $I$
is the nuclear spin; the superscript $\pi$ represents the parity.
Experimental dipole and quadrupole values are from
Ref.~\cite{PRag89}. Octupole and hexadecapole values are from a
single-particle approximation (see Appendix \ref{Apdx:spm}).
Lifetimes for the $^3P_2$ states are from Ref.~\cite{PorDer04}}
\begin{ruledtabular}
\begin{tabular}{cllllll}
\multicolumn{1}{c}{Isotope}&
\multicolumn{1}{c}{$I^{\pi}$}&
\multicolumn{1}{c}{$\mu^{\text{exp.}}$ ($\mu_N$)}&
\multicolumn{1}{c}{$Q^{\text{exp.}}$ (b)}&
\multicolumn{1}{c}{$\Omega^{\text{s.p.}}$ ($\text{b}\times\mu_N$)}&
\multicolumn{1}{c}{$\Pi^{\text{s.p.}}$ ($\text{b}^2$)}&
\multicolumn{1}{c}{$^3P_2$ Lifetime (s)}\\
\hline
$^{9 }$Be & $3/2^-$ & $-$1.177492(17)  & $+$0.053(3)  & $-$0.073 & 0.00 & \\
$^{25}$Mg & $5/2^+$ & $-$0.85545(8)    & $+$0.201(3)  & $-$0.15  & 0.00 & $1.06\times10^3$\\
$^{43}$Ca & $7/2^-$ & $-$1.317643(7)   & $-$0.049(5)  & $-$0.23  & 0.00 & $5.13\times10^2$\\
$^{87}$Sr & $9/2^+$ & $-$1.0936030(13) & $+$0.335(20) & $-$0.38  & 0.00 & $1.29\times10^2$\\
\end{tabular}
\end{ruledtabular}
\end{table*}

$^3P_2$ states in alkaline-earth atoms are well suited for extracting
nuclear magnetic octupole moments. We list the relevant isotopes in
Table~\ref{Tab:Isotopes}. The atomic lifetimes are longer than 100
seconds and are long enough to allow for a sufficiently high spectral
resolution. Moreover, successful trapping of the $^3P_2$
alkaline-earth atoms Sr and Ca has been recently reported in a
number of laboratories and there are ongoing efforts with
Mg~\cite{GruHem02,NagSimLah03,XuLofHal03,YasKat04}.
A typical trapping time is longer than 10 seconds. Since the
magnetic trap perturbs the atomic level structure, a plausible
experiment could involve microwave spectroscopy of freely-falling
atoms when trapping fields are turned off. Even then, the
interrogation time is sufficiently long to permit a determination
of the magnetic octupole HFS constant. %~\cite{KilPrivate07}.
% Tom Killian: 20 ms observation time
To extract the nuclear octupole moment from the measured HFS
constant, one requires electronic structure calculations. As shown
in this paper, such supporting calculations can be carried out with
an accuracy of a few per cent.

This paper is organized as follows. In Section~\ref{Sec:HFI} we
recapitulate the major results from the relativistic theory of the
hyperfine interaction. The consideration includes first-order
effects of nuclear dipole, quadrupole, octupole, and hexadecapole
moments and second-order dipole-dipole and dipole-quadrupole effects.
In Section~\ref{Sec:CIMBPT}, we focus on the electronic structure
aspects of the problem and evaluate electronic coupling constants
using a configuration-interaction (CI) method coupled with
many-body perturbation theory (MBPT). We refer to this
approximation as the CI+MBPT method. Further, in
Section~\ref{Sec:Sr87}, we discuss our results for the HFS
of the $^3\!P_2$ level in $^{87}$Sr. Results for
isotopes of Be, Mg, and Ca  are tabulated in
Appendix~\ref{Apdx:OtherIsotopes}. Finally, conclusions are drawn
in Section~\ref{Sec:Discussion}. The paper also has several
appendices where we tabulate results of a technical nature.
Unless specified otherwise, atomic units, $\hbar=|e|=m_e=1$, and
Gaussian electromagnetic units are employed throughout.

\section{Nuclear moments and the hyperfine interaction}
\label{Sec:HFI}
 The electric and vector potential of a nucleus
modeled as a collection of point-like multipole moments may be
expressed as
\begin{align}
\varphi\left(  \mathbf{r}\right)& =\sum_{k,\mu}\left(  -1\right)  ^{\mu}%
\frac{1}{r^{k+1}}C_{k,\mu}\left(  \mathbf{\hat{r}}\right)  \mathcal{Q}%
_{k,-\mu}, \nonumber
\\
\mathbf{A(r)} & =-i\sum_{k,\mu}\left(  -1\right)  ^{\mu}\frac{1}{r^{k+1}}%
\sqrt{\frac{k+1}{k}}\mathbf{C}_{k,\mu}^{(0)}\left(
\mathbf{\hat{r}}\right) \mathcal{M}_{k,-\mu},\label{MPexp}
\end{align}
where $C_{k,\mu}$ and $\mathbf{C}_{k,\mu}^{(0)}$ are normalized
spherical harmonics and vector spherical harmonics, respectively
(see, for example, Ref.~\cite{Joh07}). In these expressions,
$\mathcal{Q}_{k,\mu}$ and $\mathcal{M}_{k,\mu}$ are components of
the nuclear electric and magnetic 2$^{k}$-pole (MJ and EJ) moment
operators, respectively. Each of the moments is an irreducible
tensor operator of rank $k$. Parity and time-reversal symmetries
require that $k$ is even for the electric moments and $k$ is odd for
the magnetic moments. Components of these tensors are conventionally
parameterized using c-numbers that are defined as expectation values
of the zero components of the operators in a nuclear stretched
state$~|I,M_{I}=I\rangle$. We will employ the following notation for
the \textquotedblleft stretched\textquotedblright\ matrix element of
a tensor operator $O_{k,\mu}$:
\[
\langle O_{k}\rangle_{I}\equiv\langle
I,M_{I}=I|O_{k,0}|I,M_{I}=I\rangle.
\]
In particular, the magnetic-dipole, electric-quadrupole,
magnetic-octupole, and electric-hexadecapole moments of the nucleus are defined as
\begin{equation*}
\begin{array}{ccr}
\mu &=&\langle\mathcal{M}_{1}\rangle_{I},\\
Q   &=&2\langle\mathcal{Q}_{2}\rangle_{I},\\
\Omega &=&-\langle\mathcal{M}_{3}\rangle_{I},\\
\Pi &=&\langle\mathcal{Q}_{4}\rangle_{I}.
\end{array}
\end{equation*}
The stretched matrix elements are related to the reduced matrix
elements by
\[
\langle I||O_{k}||I\rangle\equiv~\left(
\begin{array}
[c]{ccc}%
I & k & I\\
-I & 0 & I
\end{array}
\right)  ^{-1}\langle O_{k}\rangle_{I}.%
\]

The interaction Hamiltonian for a single electron in an
electromagnetic field is given by
\[
h_\text{em}\left(\mathbf{r}\right) =\mathbf{\alpha}\cdot\mathbf{A}\left(
\mathbf{r}\right) -\varphi\left(  \mathbf{r}\right),
\]
and the total electromagnetic interaction Hamiltonian is then given
by
\[
H_\text{em} =\sum_i h_\text{em}\left(\mathbf{r}_i\right),
\]
where the sum runs over all electron coordinates. Substituting the
multipolar expansions (\ref{MPexp}) into the electromagnetic
interaction Hamiltonian, we arrive at an expression for
rotationally-invariant hyperfine-interaction (HFI) Hamiltonian in
the form of
\[
H_\text{HFI}=\sum_{k,\mu}\left(  -1\right)
^{\mu}T_{k,\mu}^{e}T_{k,-\mu}^{n}.
\]
Here the spherical tensors (of rank $k$) $T_{k,\mu}^{e}$ act on
electronic coordinates and $T_{k,\mu}^{n}$ operate in the nuclear
space. We identify
$T_{k,\mu}^{n}\equiv\mathcal{M}_{k,\mu}$ for odd $k$ and $T_{k,\mu}^{n}%
\equiv\mathcal{Q}_{k,\mu}$ for even $k$. Explicitly, electronic
tensors read
\begin{equation*}
T_{k,\mu}^{e} = \sum_i t_{k,\mu}^{e}\left(\mathbf{r}_i\right),
\end{equation*}
with
\begin{equation}
t_{k,\mu}^{e}\left(\mathbf{r}\right)=\left\{
\begin{array}{c l}
-\frac{1}{r^{k+1}}C_{k,\mu}\left(  \mathbf{\hat{r}}\right)
& \quad\mbox{electric (even $k$),} \\[1ex]
-\frac{i}{r^{k+1}}\sqrt{\frac{k+1}{k}
}\mathbf{\alpha}\cdot\mathbf{C}_{k,\mu}^{(0)}\left(\mathbf{\hat{r}}\right)
& \quad\mbox{magnetic (odd $k$).} \\
\end{array}\right.
\label{Eq:TeGeneral}
\end{equation}
Matrix elements of these operators for Dirac bi-spinors are listed
in Appendix~\ref{Apdx:Mels}.

Now we recapitulate the application of perturbation theory to
determining the hyperfine structure of atomic levels. In the
presence of the multipolar fields produced by the nucleus, the total
electronic angular momentum $\mathbf{J}$ is no longer conserved. The
conserved angular momentum includes the nuclear angular momentum
$\mathbf{I}$; explicitly this conserved angular momentum is
$\mathbf{F}=\mathbf{I} +\mathbf{J}$. The proper basis consists of
states $|\gamma IJFM_{F}\rangle$  generated by coupling nuclear and
electronic wave functions,
\[
|\gamma
IJFM_{F}\rangle=\sum_{M_{J},M_{I}}C_{JM_{J};IM_{I}}^{FM_{F}}\left\vert
\gamma JM_{J}\right\rangle \left\vert IM_{I}\right\rangle ,
\]
with $\gamma$ encapsulating remaining electronic quantum numbers and
the coupling coefficients being the conventional Clebsch-Gordon
coefficients. For fixed values of $J$ and $I$, the values of $F$
range in $|J-I|\leq F\leq J+I$, implying that an unperturbed level
with angular momentum $J$ is split into $2J+1$ levels for $J<I$
and into $2I+1$ levels otherwise.

A matrix element of the HFI in the coupled basis is
\begin{align}
\langle\gamma^{\prime}IJ^{\prime}F^{\prime}M_{F}^{\prime}|H_\text{HFI}|\gamma
IJFM_{F}\rangle =\delta_{F^{\prime}F}\delta_{M_{F}^{\prime}M_{F}}
\qquad\qquad\nonumber\\
\times(-1)^{I+J+F} \sum_{k}\left\{
\begin{array}
[c]{ccc}%
F & J & I\\
k & I & J^{\prime}%
\end{array}
\right\}  \,\langle\gamma^{\prime}J^{\prime}||T_{k}^{e}||\gamma J%
\rangle\langle I||T_{k}^{n}||I\rangle,\label{Eq:HFIgen}
\end{align}
the $\delta$-symbols reflecting the scalar character of the HFI in
the coupled basis. Hyperfine corrections to an unperturbed level
are given to first order by the diagonal matrix elements of $H_\text{HFI}$.
For convenience, we
write this correction in terms of the $F$-independent product of the
stretched matrix elements
\begin{eqnarray*}
W^{(1)}_{F}&=&\langle\gamma IJFM_{F}|H_\text{HFI}|\gamma IJFM_{F}\rangle \\
         &=&\sum_{k} X_{k}(IJF)~\langle
            T_{k}^{e}\rangle_{J}\,\langle T_{k}^{n}\rangle_{I},
\end{eqnarray*}
with%
\begin{eqnarray*}
X_{k}(IJF)=(-1)^{I+J+F} \frac{ \left\{
\begin{array}
[c]{ccc}%
F & J & I\\
k & I & J
\end{array}
\right\}} {\left(
\begin{array}
[c]{ccc}%
J & k & J\\
-J & 0 & J
\end{array}
\right)\left(
\begin{array}
[c]{ccc}%
I & k & I\\
-I & 0 & I
\end{array}
\right)}.
\end{eqnarray*}
The first-order $F$-dependent effects are conventionally
parameterized in terms of the hyperfine structure constants
$A,B,C,D,\cdots$. Successive labels correspond to contributions of a
multipole of increasing multipolarity, e.g., $A$ is due to the
magnetic dipole moment, $B$ due to the electric quadrupole moment,
etc. The constants, up to $D$, are defined as in~\cite{DanFerGeb74}:
\begin{equation}
\label{Eq:HFconsts}
\begin{array}{rrrrr}
A &=& \frac{1}{IJ}\cdot
      \langle T_{1}^{n}\rangle_{I}\langle T_{1}^{e}\rangle_{J}
  &=&\frac{1}{IJ}\cdot\mu\langle T_{1}^{e}\rangle_{J},\\
B &=& 4\cdot\langle T_{2}^{n}\rangle_{I}\langle T_{2}^{e}\rangle_{J}
  &=&2\cdot Q\langle T_{2}^{e}\rangle_{J},\\
C &=&\langle T_{3}^{n}\rangle_{I}\langle T_{3}^{e}\rangle_{J}
  &=&-\Omega\langle T_{3}^{e}\rangle_{J},\\
D &=&\langle T_{4}^{n}\rangle_{I}\langle T_{4}^{e}\rangle_{J}
  &=& \Pi\langle T_{4}^{e}\rangle_{J}.
\end{array}
\end{equation}

The second-order (in the HFI) correction for the state described by electronic
quantum numbers $\gamma$ and $J$ is
\begin{equation*}
W^{(2)}_{F} = \sum_{\gamma^{\prime}J^{\prime}} \frac{\left\vert
\langle\gamma^{\prime}IJ^{\prime}FM_{F}|H_\text{HFI}|\gamma
IJFM_{F}\rangle \right\vert ^{2}}{E_{\gamma
J}-E_{\gamma^{\prime}J^{\prime}}},
\end{equation*}
where the sum excludes the case
$\left(\gamma^{\prime}J^{\prime}\right)=\left(\gamma J\right)$. With
Eq.~(\ref{Eq:HFIgen}) this can be expressed explicitly in terms of
reduced matrix elements as
\begin{align*}
W^{(2)}_{F} =\sum_{\gamma^{\prime}J^{\prime}} \frac{1}{E_{\gamma
J}-E_{\gamma^{\prime}J^{\prime}}} \sum_{k_1,k_2} \left\{
\begin{array}{ccc}
 F   & J & I \\
 k_1 & I & J^{\prime}
\end{array}
\right\} \left\{
\begin{array}{ccc}
 F   & J & I \\
 k_2 & I & J^{\prime}
\end{array}
\right\}\\
\times\langle\gamma^{\prime}J^{\prime}||T_{k_1}^{e}||\gamma J\rangle
\langle\gamma^{\prime}J^{\prime}||T_{k_2}^{e}||\gamma J\rangle
\langle I||T_{k_1}^{n}||I\rangle \langle I||T_{k_2}^{n}||I\rangle.
\end{align*}
For the case of interest, where $J=2$, the sum over ($\gamma'\, J'$) is
dominated by contributions from the adjacent fine structure levels ($\gamma\, 1$)
and the sum over $k_1 k_2$ is dominated by the dipole-dipole
$\left(k_1=k_2=1\right)$ term. If we limit ourselves to the
dipole-dipole and dipole-quadrupole terms,
then we may express the second-order correction as
\begin{multline*}
W^{(2)}_{F} = \left|\left\{
\begin{array}{ccc}
 F & J & I \\
 1 & I & J-1
\end{array}
\right\}\right|^2 \eta \\
+ \left\{
\begin{array}{ccc}
 F & J & I \\
 1 & I & J-1
\end{array}
\right\}
\left\{
\begin{array}{ccc}
 F & J & I \\
 2 & I & J-1
\end{array}
\right\} \zeta
\end{multline*}
where $\eta$ and $\zeta$ are $F$-independent terms given by
\begin{align}
\label{Eq:eta} \eta & = \frac{(I+1)(2I+1)}{I}\,\mu^2\, \frac{
\left| \langle\gamma J-1\|T_{1}^{e}\|\gamma J\rangle\right|^{2} }
{E_{\gamma J}-E_{\gamma J-1} }\\
\zeta &= \frac{(I+1)(2I+1)}{I}\,\sqrt{\frac{2I+3}{2I-1}}
\times\nonumber \\
& \hspace{2em} \mu\, Q\, \frac{ \langle\gamma J-1\|T_{1}^{e}\|\gamma J\rangle
 \langle\gamma J-1 \|T_{2}^{e} \|\gamma J\rangle} {E_{\gamma J}-E_{\gamma J-1}}
\end{align}
In the above expressions, $\gamma$ denotes a given
fine structure term such as $^{3}P_{2}$ and that $J' = J-1 = 1$.

\section{CI+MBPT electronic wave functions}
\label{Sec:CIMBPT}
A precise analysis of the hyperfine structure
depends on accurate electronic wave functions as well as nuclear
multipole moments. To obtain accurate electronic wave functions, we
used a combined method of configuration interaction (CI) and
many-body perturbation theory (MBPT), which we refer to as CI+MBPT.
The CI+MBPT method is described in detail, for example, in
Refs.~\cite{DzuFlaKoz96,SavJoh02}.
 Here, we restrict the presentation to a qualitative
discussion and then apply the CI+MBPT method to determination of the HFS
couplings.

The accuracy of the CI method is limited only by the completeness of
the set of configurations used.  In the context of CI+MBPT we
refer to the sub-space containing the configurations to be treated
by the CI method as the model space. In CI+MBPT, additional
contributions from configurations which are not in the model space
can be accounted for by MBPT. For our purposes, we are interested in
finding accurate wave functions. It is worth noting that the
wave functions determined by the CI+MBPT analysis lie completely
within the model space.

In deciding which configurations are to be included in the model
space, we consider two things. First, configurations which interact
strongly with the configuration of interest must be included in
the model space. The main purpose here is to expand the model space
to a degree in which the wave function can be described accurately.
Second, configurations with energies which are nearly degenerate
with the configuration of interest should also be included in the
model space. These configurations lead to convergence problems in
MBPT and are treated non-perturbatively by the CI method. Of
course, ``strongly'' and ``nearly degenerate'' are relative terms
which depend on the level of accuracy desired.

For the alkaline earth systems considered in this paper, we start
with a lowest-order description of our system as two valence
electrons outside a closed core in a central field.
Configurations involving different excitations of the valence
electrons outside the closed core tend to interact strongly and have
relatively close energies. These configurations compose our model
space. Additional contributions from configurations involving
excitations from core electrons are then accounted for within the
framework of MBPT.

\begin{figure}[h]
\begin{center}
\includegraphics[scale=1]{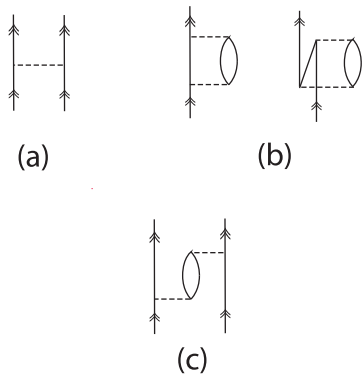}
\end{center}
\caption{Sample Brueckner-Goldstone diagrams included in the  effective CI+MBPT Hamiltonian.
(a) Coulomb interaction between the two valence electrons. The double-arrows
indicate that the single-particle states originate/terminate in the model space.
(b) Brueckner (core-polarization) diagrams. (c) Screening diagrams.
 } \label{Fig:diagrams}
\end{figure}

The relevant diagrams are shown in Fig.~\ref{Fig:diagrams}. The
Coulomb interaction between the two valence electrons is taken into
account essentially to all orders of perturbation theory and the
remaining many-body effects (involving core excitations in the
intermediate state) are treated in the second order of MBPT. In
Fig.~\ref{Fig:diagrams} the single-particle states are assumed to be
generated in the frozen-core ($V^{N-2}$) Dirac-Hartree-Fock (DHF)
approximation where the so-called one-body (loop) diagrams vanish
identically.

The specific approach that we used here is somewhat more
sophisticated~\cite{SavJoh02}: as the single-particle orbitals we
employ the so-called Brueckner orbitals. This approach  incorporates
the diagrams Fig.~\ref{Fig:diagrams}(b) from the onset and
also subsumes higher-order chains. Then the single-particle states
(vertical lines) in the remaining diagrams are replaced by the
Brueckner orbitals. The entire scheme incorporates all the
second-order MBPT effects and includes certain classes of diagrams
to all orders. Technically, generating Brueckner orbitals requires
an energy-dependent self-energy operator (see, e.g.,
Ref.~\cite{SavJoh02}), $\Sigma(\varepsilon_\kappa)$, which generally
leads to non-orthogonal basis sets. We avoid this problem by fixing
$\varepsilon_\kappa$ to the $V^{N-2}$ DHF energy of the
lowest-energy valence orbital for a given angular symmetry $\kappa$
($\kappa$ is defined explicitly in Appendix~\ref{Apdx:Mels}).

To illustrate the predictive capabilities of the CI+MBPT method, in
Table~\ref{Tab:CIMBPT-Sr} we compare the theoretical energy values with the
experimental values for Sr. The difference does not exceed a few 100
$\mathrm{cm}^{-1}$ for all the calculated energy levels. While the accuracy can
be improved further,  this level is sufficient for the goals of the present paper.

\begin{table}
\caption{\label{Tab:CIMBPT-Sr}  Energies of Sr I obtained from CI+MBPT
are compared to experimental values. The table is partitioned
into states of definite $J$ and parity. Energies are referenced from
the ground $5s^2\,^1\!S_0$ state. All
energy values are in cm$^{-1}$.}
\begin{ruledtabular}
\begin{tabular}{@{\quad}ccll@{\qquad}c@{\qquad}c@{\qquad}c@{\quad}}
 $J$  & Parity & \multicolumn{2}{c@{\qquad}}{Level} & CI+MBPT & Expt. & Diff. \\
\hline
 0    &      even & $5s6s$ & $^1S_0$ & 30766 & 30592 & 174   \\
      &           & $5p^2$ & $^3P_0$ & 35798 & 35193 & 605   \\
      &           & $5p^2$ & $^1S_0$ & 37553 & 37160 & 393   \\
      &           &        &         &       &       &       \\
 1    &      odd~ & $5s5p$ & $^3P_1$ & 14841 & 14504 & 337   \\
      &           & $5s5p$ & $^1P_1$ & 21818 & 21699 & 119   \\
      &           & $5s6p$ & $^3P_1$ & 34062 & 33868 & 194   \\
      &           & $5s6p$ & $^1P_1$ & 34293 & 34098 & 195   \\
      &           &        &         &       &       &       \\
 2    &      odd~ & $5s5p$ & $^3P_2$ & 15212 & 14899 & 313   \\
      &           & $4d5p$ & $^3F_2$ & 33836 & 33267 & 569   \\
      &           & $4d5p$ & $^1D_2$ & 34149 & 33827 & 322   \\
\end{tabular}
\end{ruledtabular}
\end{table}

Solving the CI+MBPT problem reduces to diagonalizing the effective
Hamiltonian in the model space. With the CI+MBPT wave functions we
proceed to evaluating matrix elements. Details are given in
Ref.~\cite{SavJoh02}; the method builds upon the formalism
originally developed for He-like systems~\cite{JohPlaSap95}. We
also include an important chain of diagrams of the random-phase
approximation (RPA) in the evaluation of matrix elements.
Qualitatively, RPA accounts for screening of externally-applied
fields (in our particular case these are the multipolar nuclear
fields) by the core electrons.

Numerically, the calculations were carried out using {\em B}-spline
basis sets. It is worth mentioning our modification to the original
scheme~\cite{JohBluSap88} of generating the orbital sets. In that
scheme, boundary conditions are imposed on the small and large
components of the wave function which were found to effectively
dispense of spurious states in the orbital sets (more accurately,
these spurious states were shifted towards the end of the spectrum).
The disadvantage that follows is that the resulting wave functions
are not highly accurate near the nucleus, leading to inaccuracies
in the HFI integrals of
Appendix~\ref{Apdx:Mels}. To remedy this problem, we followed the
prescription for creating a dual kinetic-balance (DKB) basis set, as
introduced by Shabaev {\em et. al.}~\cite{ShaTupYer04}. The DKB
basis set is devoid of the problem of spurious states and
capable of accurately representing the wave functions near the
nucleus. The specific formulas used here for generating DHF orbital
sets with a DKB basis set derived from {\em B}-splines are presented
in Ref.~\cite{BelDer07}.

Another technical issue is the appearance of  ``intruder'' states
in screening diagrams, Fig.~\ref{Fig:diagrams}(c).
Intermediate states (the diagram is cut across horizontally) there involve
core-excited states. Since our model (CI) space is essentially complete,
such core-excited states become embedded in the two-particle energy spectrum of
the lowest-order model Hamiltonian. This leads to singular energy denominators
in the diagrams  Fig.~\ref{Fig:diagrams}(c). We remedied this problem
by evaluating screening corrections only to the two particle states that had energies
below the lowest-energy core excitation. For a typical basis set the resulting
subspace involved roughly 10\% of the entire model space. Arguably, this prescription
recovers most of the relevant correction since this low-energy
subspace contains the dominant configurations.

\section{Hyperfine structure of $^{87}$Sr}
\label{Sec:Sr87}%
Using the techniques described in the previous sections, we consider as
a specific example the hyperfine structure of the lowest-energy
$^{3}P_{2}$ level of $^{87}$Sr.

\subsection{Extracting the magnetic-octupole constant from measurement of
hyperfine intervals}
The nuclear spin of the stable
$^{87}$Sr isotope is $I=9/2$, and so there are five hyperfine
structure levels $F=5/2,\cdots,13/2$. First-order
corrections are then characterized by four hyperfine structure
constants $A, B, C,$ and $D$. In addition, the second-order dipole-dipole
interaction, characterized by $\eta$
and the dipole-quadrupole interaction, characterized by $\zeta$,
 mix the $^{3}P_{2}$ state with the nearby fine
structure $^{3}P_{1}$ state.  Using the expressions given in
Section~\ref{Sec:HFI}, we may write the hyperfine correction for
$^{87}$Sr in terms of these constants as
\begin{align*}
W_{5/2} &= -11A+\frac{11}{24}B-\frac{143}{42}C
               +\frac{143}{18}D,\\
W_{7/2} &=
-\frac{15}{2}A+\frac{1}{48}B+\frac{65}{21}C
-\frac{52}{3}D\\
 &\hspace{8em} +\frac{7\eta}{900} -\frac{7 \zeta}{300\sqrt{6}},\\
W_{9/2} &=
-3A-\frac{7}{24}B+\frac{13}{6}C
+\frac{91}{6}D\\
&\hspace{8em}  +\frac{32\eta}{2475} -\frac{4 \zeta}{825} \sqrt{\frac{2}{3}},\\
W_{11/2} &=  \frac{5}{2}A-\frac{7}{24}B-\frac{10}{3}C
-\frac{56}{9}D \\
&\hspace{8em}  + \frac{13\eta}{1100} +\frac{13\zeta}{550\sqrt{6}},\\
W_{13/2} &=  9A+\frac{1}{4}B+C+D.
\end{align*}
The resulting HFS intervals $\delta W_{F} = W_F - W_{F+1}$ are given in terms of the hyperfine constants as
\begin{align}
\delta W_{5/2} &=
    -\frac{7}{2}A+\frac{7}{16}B-\frac{13}{2}C
    +\frac{455}{18}D\nonumber\\
  &\hspace{9em} -\frac{7\eta}{900} +\frac{7\zeta}{300\sqrt{6}}, \nonumber\\
\delta W_{7/2} &=
    -\frac{9}{2}A+\frac{5}{16}B+\frac{13}{14}C
    -\frac{65}{2}D\nonumber \\
  &\hspace{9em} -\frac{17\eta}{3300}-\frac{\zeta}{220}\sqrt{\frac{3}{2}},\nonumber\\
\delta W_{9/2} &=
    -\frac{11}{2}A+\frac{11}{2}C
    +\frac{385}{18}D\nonumber\\
 &\hspace{9em} +\frac{\eta}{900}-\frac{\zeta}{30\sqrt{6}},\nonumber\\
\delta W_{11/2} &=
    -\frac{13}{2}A-\frac{13}{24}B-\frac{13}{3}C
    -\frac{65}{9}D\nonumber\\
 &\hspace{9em}+\frac{13\eta}{1100}   + \frac{13\zeta}{550\sqrt{6}}.
\label{Eq:dW-Sr}
\end{align}
Similar expressions for the hyperfine intervals in $^{3}P_{2}$ states of
stable isotopes of other alkaline-earth atoms are given in Appendix
\ref{Apdx:OtherIsotopes}.
Solving Eqs.(\ref{Eq:dW-Sr}) for the hyperfine constants $C$ and $D$, one finds
\begin{align}
C  & =   -\frac{  3}{  50}\delta W_{ 5/2}
      +\frac{  7}{ 550}\delta W_{ 7/2} \nonumber\\
&      +\frac{ 21}{ 275}\delta W_{ 9/2}
      -\frac{147}{3575}\delta W_{11/2}
      +\frac{7\zeta}{1375\sqrt{6}} \label{Eq:C-Sr} \\
D & =  \frac{  3}{ 350}\delta W_{ 5/2}
    -\frac{  9}{ 550}\delta W_{ 7/2} \nonumber \\
& \hspace{5em}   +\frac{  3}{ 275}\delta W_{ 9/2}
    -\frac{  9}{3575}\delta W_{11/2} \label{Eq:D-Sr}
\end{align}
Expressions for the HFS constants $C$ and $D$ in terms of
hyperfine intervals for  $^{3}P_{2}$ states in other isotopes of
the alkaline-earth atoms are given in
Appendix~\ref{Apdx:OtherIsotopes}.
The constant $C$ depends only on the second-order
dipole-quadrupole interference term $\zeta$ while
$D$ is independent of both $\eta$ and $\zeta$.
This proposition is independent of nuclear spin $I$,
as shown in Appendix~\ref{Apdx:Proof}.

To reiterate, measuring hyperfine intervals of the $^3P_2$
level should allow one to deduce
the magnetic-octupole HFS constant $C$, limited only by the knowledge of $\zeta$.
With the aid of calculations of the electronic structure factor presented below
one may extract the nuclear magnetic octupole moment of interest.

\subsection{Electronic structure factors}
From Eqs.~(\ref{Eq:HFconsts}) and (\ref{Eq:eta}) we see that the
HFS constants may be written in terms of products of electronic
matrix elements and nuclear multipole moments. We evaluate the
electronic matrix elements using the relativistic many-body method  described
in Section~\ref{Sec:CIMBPT}. We present the results
of our calculations at various levels of approximation in Table \ref{Tab:MBPTHFS}.

In generating the DHF orbitals we included nuclear-size effects by
assuming a Fermi charge distribution inside the nucleus. For
computation of HFI integrals, however, we assumed a point-size
nucleus. The observed effect on the choice of a particular model of
nuclear distribution is below our theoretical error for solving the
electronic-structure problem.

There are several observations to be made with respect to the
many-body calculations. First of all, we find that for Sr,  the
configuration $5p_{3/2} 5s_{1/2}$ provides the dominant (92\%)
contribution to the CI wave function of the $^3P_2$ state. In
particular, due to the angular selection rules, the constants $B$
and $C$ are determined by the matrix elements involving the
$5p_{3/2}$ orbital from this dominant configuration. By contrast,
the dominant $5p_{3/2} 5s_{1/2}$ configuration does not contribute
to the electric-hexadecapole (E4) constant due to selection rules
for single-particle matrix elements (see Appendix). Therefore the
E4 electronic factor is strongly suppressed, as its value is
accumulated entirely due to admixed configurations.

We find that the BO (core-polarization) corrections universally
increase the absolute value of all the constants. Qualitatively,
core polarization describes an attraction of the valence electron by
the core. This attraction leads to enhanced density closer to the
nucleus and simultaneously larger hyperfine constants. Similarly,
the computed RPA corrections show that the internal nuclear fields
are enhanced by virtual core excitations. The screening diagrams,
Fig.~\ref{Fig:diagrams}(c), qualitatively represent an effect of
``cross-talking'' between electrons via core polarization: a valence
electron polarizes the core and this induced polarization attracts
or repels another valence electron. We see for the magnetic-dipole
HFS constant that this effect is relatively weak compared to the
other many-body corrections; however, its effect is more substantial
for the electric-quadrupole and magnetic-octupole HFS constants. In
all three of these cases, the screening contribution has the effect
of decreasing the absolute value of the HFS constants.
% Comment here or in next paragraph on the screening effect on D

%On average this effect is much weaker than the direct
%core-polarization attraction of the electron (BO correction). For
%example, we find that for the magnetic-dipole HFS constant it
%contributes only 0.2\%. At the same time calculations of these
%corrections are computationally expensive and require treating
%``intruder states'', as described in Section~\ref{Sec:CIMBPT}.
%Because of these considerations we did not include the screening
%diagrams in our final tabulation.

Finally,  in Table \ref{Tab:MBPTHFS} we compare our {\em ab initio}
results with experimental values for $A$ and $B$. We find an 8\%
agreement for both constants. We believe that these accuracies are
indicative of the theoretical error for the electronic factor
entering the magnetic-octupole constant $C$. The accuracy of
computing the electronic factor for the HFS constant $D$ is
expected to be worse because the entire value is accumulated due to
correlation effects.

\begin{table*}[ht]
\caption{\label{Tab:MBPTHFS}  Breakdown of many-body corrections
to hyperfine
structure constants of $^{87}$Sr  $5s5p\,^3\!P_2$ state.
We used $\mu =-1.0936 \mu_N $ and $Q=0.305(2)$ b (Ref.~\cite{Sah06})
in tabulating $A$ and $B$ constants.
The final result is compared with experimental values from
Ref.~\cite{HeiBri77}. CI-DHF corresponds to CI values computed using single-particle
basis generated in the frozen-core ($V^{N-2}$) DHF potential.}
\begin{ruledtabular}
\begin{tabular}{ldddd}
&
\multicolumn{1}{c}{$A$, MHz} &
\multicolumn{1}{c}{$B$, MHz }  &
\multicolumn{1}{c}{$C/\Omega$, $\mathrm{MHz}/(\mu_N \times \text{b})$} &
\multicolumn{1}{c}{$D/\Pi$,  $\mathrm{MHz}/\text{b}^2$ } \\
\hline
CI-DHF             &   -147.1    & 35.6      &  3.54  \times 10^{-4} &  0.54   \times 10^{-12} \\
&\multicolumn{4}{c}{   Many-body corrections }\\
$\Delta$ BO        &   -41.9     & 10.2      &  1.03  \times 10^{-4} &  0.85   \times 10^{-12} \\
$\Delta$ Screen    &     0.4     & -4.6      & -0.42  \times 10^{-4} &  2.71   \times 10^{-12} \\
$\Delta$ RPA       &   -42.0     & 21.0      &  1.15  \times 10^{-4} &  0.55   \times 10^{-12} \\
\hline
Final              & -230.6      & 62.2      &  5.30  \times 10^{-4} &  4.65   \times 10^{-12}   \\
Experiment         & -212.765(1) & 67.215(15)&  &  \\
\end{tabular}
\end{ruledtabular}
\end{table*}
We have carried out similar many-body calculations for the
parameters $\eta$ and $\zeta$ entering the second-order correction
to the hyperfine constants. For $^{87}$Sr, we find $\eta = 6.65 \
\text{MHz}$ and $\zeta = 0.529\ \text{MHz}$.  These second-order
corrections are scaled to the experimental $A$ and $B$ coefficients and are accurate to
2\%.  The second-order dipole-quadrupole contribution to $C$ is
\begin{equation}
\Delta C(^{87}\mathrm{Sr})   = \frac{7\zeta}{1375\sqrt{6}} = 1.10(2)\, \text{kHz} .
\label{Eq:dCnumSr}
\end{equation}

In this section we have described our calculations for $^{87}$Sr. The corresponding
results for $^{9 }$Be, $^{25}$Mg, and $^{43}$Ca are given in Appendix
\ref{Apdx:OtherIsotopes}.

\section{Discussion}
\label{Sec:Discussion}

At this point we combine the computed electronic structure
factors for the magnetic-octupole constants
(Table~\ref{Tab:MBPTHFS} and Table~\ref{Tab:others}) and the
nuclear shell-model prediction for the M3 moment $\Omega$ (Table
~\ref{Tab:Isotopes} and Appendix~\ref{Apdx:spm}). We find that
\begin{eqnarray}
C(^{9} \mathrm{Be})  &=& -3.57\times 10^{-2}  \,\mathrm{Hz} \,, \nonumber \\
C(^{25} \mathrm{Mg}) &=& -2.57  \,\mathrm{Hz} \,, \label{Eq:Cest} \\
C(^{43}\mathrm{Ca})  &=& -16.2  \,\mathrm{Hz} \,, \nonumber\\
C(^{87}\mathrm{Sr})  &=& -201 \, \mathrm{Hz} \,. \nonumber
\end{eqnarray}
 In particular for Sr,
using Eq.(\ref{Eq:C-Sr}), we may deduce that an experimental
sensitivity in measuring the HFS intervals on the order of
$\sigma_{\delta W}\approx 1$ kHz would result in an uncertainty in
the $C$ constant on the order of $\sigma_C \approx
0.11~\sigma_{\delta W}\approx 100$ Hz and would thus be capable of
revealing the effects of a $C$ constant of the predicted
magnitude.

The expression for the constant of interest $C$ in term of HFS splittings
contains the second-order dipole-quadrupole correction, see, e.g., Eq.(\ref{Eq:C-Sr}) 
for Sr. 
If the experiment measures HFS intervals with
a vanishingly small error bar, the extraction of $\Omega$ would be limited
by the error in this correction. Our estimated error bar for Sr, Eq.~(\ref{Eq:dCnumSr})
is 20 Hz, which translated into 10\% error bar for  $C(^{87}\mathrm{Sr})$ of
predicted magnitude. Similar conclusion holds for Ca, while for Mg our 
estimated uncertainty in the dipole-quadrupole correction is comparable with
the predicted size of $C$.
For Be, our present uncertainty of 0.2 Hz 
in the dipole-quadrupole correction precludes clean extraction of the octupole
moment. 
Notice, however, that  the nuclear shell model estimates of the nuclear octupole moment 
may be unreliable.
If the $^{133}$Cs experiment~\cite{GerDerTan03} is of any indication, 
the ``true'' size of the
octupole constant may be much larger (factor of 40) than predicted. Then
the dipole-quadrupole corrections become mostly irrelevant.

We emphasize that the values~(\ref{Eq:Cest})  are only estimates based on the nuclear-shell model;
measuring $C$ would show deviations from these estimates. In a particular case
of $^{133}\mathrm{Cs}$ the measured and the predicted values differed by a factor of 40~\cite{GerDerTan03}. It
remains to be seen if such large deviations from the nuclear shell model would be
revealed experimentally for the nuclei considered in the present work.

\acknowledgements
The work of KB and AD was supported in part by National Science Foundation
grant No. PHY-06-53392 and the work of WRJ was supported in part by NSF grant No.
PHY-04-56828.

\appendix

\section{Nuclear Moments from a single particle model}
\label{Apdx:spm} A crude approximation to the nuclear moments can be
achieved by representing the nucleus by a single nucleon. For
even-odd (even number of protons, odd number of neutrons) nuclei we
use a single neutron, and for odd-even nuclei we use a single
proton. Using formulas by \citet{Sch55}, we may write the moments in
this single particle model as
\begin{eqnarray*}
\mu^{\text{s.p.}} &=& \mu_N I\times\
\begin{cases}
[g_l +(g_s-g_l)/2I]     & \text{for $I=l+\frac{1}{2}$}, \\
[g_l -(g_s-g_l)/(2I+2)] & \text{for $I=l-\frac{1}{2}$},
\end{cases}\\
Q^{\text{s.p.}} &=& -e\langle r^2\rangle g_l\frac{2I-1}{2I+2}, \\
\Omega^{\text{s.p.}} &=& \mu_N\langle r^2\rangle
\frac{3}{2}\frac{(2I-1)}{(2I+4)(2I+2)} \\ &&\times
\begin{cases}
(I+2)[(I-\frac{3}{2})g_l + g_s] & \text{for $I=l+\frac{1}{2}$}, \\
(I-1)[(I+\frac{5}{2})g_l - g_s] & \text{for $I=l-\frac{1}{2}$}, \\
\end{cases}\\
\Pi^{\text{s.p.}} &=& -e\langle r^2\rangle g_l\frac{3}{8}
\frac{(2I-1)(2I-3)}{(2I+4)(2I+2)},
\end{eqnarray*}
where $g_l = +1$, $g_s = 5.58$ for a proton and $g_l = 0$, $g_s =
-3.83$ for a neutron. All of the stable isotopes considered in this
paper have even-odd nuclei. This has the immediate consequence
$Q^{\text{s.p.}}=0$ and $\Pi^{\text{s.p.}}=0$ for all isotopes.
Furthermore, an examination of the momentum $I$ and parity $\pi$ in
Table \ref{Tab:Isotopes} reveals that the nucleon for each isotope
must have an orbital momentum $l$ satisfying $I=l+1/2$. With this
additional consideration, $\mu^{\text{s.p.}}=-1.92\mu_N$ for all
isotopes, and the expression for the octuple moment is reduced to
\begin{equation*}
\Omega^{\text{s.p.}} = \mu_N\langle r^2\rangle g_s
\frac{3}{4}\frac{(2I-1)}{(2I+2)}.
\end{equation*}
Approximating the root-mean-square value of the nuclear radii, $\langle
r^2\rangle^{1/2}$, as (in units of fm) 2.52, 3.05, 3.48, and 4.24
for the cases of $^{9 }$Be, $^{25}$Mg, $^{43}$Ca, and $^{87}$Sr,
respectively, yields the values for $\Omega^{\text{s.p.}}$ given in
Table \ref{Tab:Isotopes}.

In the single particle model the electromagnetic moments of the
nuclei are given by the appropriate expectation values for the
valence nucleon shell. In this model the even-odd nuclei would have
electromagnetic moments determined by the valence neutron. In
particular, since the neutron doesn't have an electric charge, all
electric moments vanish. Certainly, the observed nonzero Q-moment
provides an information on such quantities as nuclear deformation.
Similarly, an observation of the electric hexadecapole moment, which
is zero in the single particle approximation, should provide
similar information on the nuclear distortion.

\section{Matrix elements of the electronic tensor operator}
\label{Apdx:Mels}

Here we compile expressions for the matrix
elements of the single-particle electronic HFI coupling operators
$t_{k,\mu}^{e}\left(\mathbf{r}\right)$ given in Eq.~(\ref{Eq:TeGeneral}).
We use the conventional parametrization of the Dirac bi-spinors,
\begin{equation*}
|n\kappa m\rangle=\frac{1}{r}\left(
\begin{array}
[c]{c}%
iP_{n\kappa}(r)\ \Omega_{\kappa m}(\hat{r})\\
Q_{n\kappa}(r)\ \Omega_{-\kappa m}(\hat{r})
\end{array}
\right),  \label{Eq:BiSpinorSpher}%
\end{equation*}
where $\kappa =(l-j)\left( 2j+1\right) $ and $\Omega$ are spinor
functions. With this parametrization,
we find that the reduced matrix elements for the HFI couplings to
electric moments of the nucleus are given by%
\begin{eqnarray*}
\langle n^{\prime}\kappa^{\prime}||t_{k}^{e}||n\kappa\rangle &=
&-\langle\kappa^{\prime}||C_{k}||\kappa\rangle \\
&&\times\int_{0}^{\infty}\frac{dr}{r^{k+1}}\left(
P_{n^{\prime}\kappa^{\prime}}P_{n\kappa}+Q_{n^{\prime}\kappa^{\prime}}Q_{n\kappa%
}\right)  ,
\end{eqnarray*}
and for couplings to magnetic moments these are%
\begin{eqnarray*}
\langle n^{\prime}\kappa^{\prime}||t_{k}^{e}||n\kappa\rangle &=
&\langle\kappa^{\prime}||C_{k}||-\kappa\rangle
\left(\frac{\kappa^{\prime}+\kappa}{k}\right) \\
&&\times\int_{0}^{\infty}\frac{dr}{r^{k+1}}\left(  P_{n^{\prime}\kappa^{\prime}}Q_{n%
\kappa}+Q_{n^{\prime}\kappa^{\prime}}P_{n\kappa}\right)  .
\end{eqnarray*}
Selection rules for these matrix elements follow from those for the
matrix elements of the $C$-tensor: $\left\vert j-j'\right\vert \leq
k\leq j+j'$ and the sum $l+l'$ must be even.

\section{Hyperfine structure of the $^3P_2$ state for $^{9 }$Be, $^{25}$Mg, and $^{43}$Ca}
\label{Apdx:OtherIsotopes} This appendix contains a compilation of
expressions relating the hyperfine  intervals and the
hyperfine structure constants for the $^{3}P_{2}$ states of $^{9
}$Be, $^{25}$Mg, and $^{43}$Ca. Computed HFS constants for
these isotopes, given in Table~\ref{Tab:others}, are also
included.

$^{9 }$Be, $I=3/2$:
\begin{eqnarray*}
\delta W_{1/2} &=&
    -\frac{3}{2}A+\frac{7}{8}B-28C
    -\frac{11\eta}{600} +\frac{\sqrt{3}\zeta}{200},   \nonumber\\
\delta W_{3/2} &=&
    -\frac{5}{2}A+\frac{5}{8}B+20C
    -\frac{\eta}{120} -\frac{\sqrt{3}\zeta}{40},  \nonumber\\
\delta W_{5/2} &=&
    -\frac{7}{2}A-\frac{7}{8}B-7C
    +\frac{7\eta}{200}+\frac{7\zeta}{200\sqrt{3}},
\end{eqnarray*}
\[
C = -\frac{  1}{  50}\delta W_{ 1/2 }
      +\frac{  1}{  50}\delta W_{ 3/2}
    -\frac{  1}{ 175}\delta W_{ 5/2 }+\frac{\zeta}{500\sqrt{3}} \label{Eq:C-Be}
\]
For $^{9}$Be, we find that $\eta = 25.09(4)$~MHz and $\zeta=0.2939(2)$~MHz,
leading to a value $\Delta C = 0.3394(2)$~kHz for the second-order correction to $C$.

$^{25}$Mg, $I=5/2$:
\begin{eqnarray*}
\lefteqn{\delta W_{1/2} =   -\frac{3}{2}A+\frac{9}{20}B-\frac{54}{5}C +90D }\hspace{11em}\\
 &&   -\frac{\eta}{100} +\frac{\zeta}{50\sqrt{2}},   \\
\lefteqn{\delta W_{3/2 } = -\frac{5}{2}A+\frac{1}{2}B-3C -75D}\hspace{11em}\\
 &&   -\frac{13\eta}{1260} -\frac{\zeta}{210\sqrt{2}},  \\
\lefteqn{\delta W_{5/2} = -\frac{7}{2}A+\frac{7}{40}B+\frac{49}{5}C +35D}\hspace{11em}\\
 && -\frac{\eta}{900} -\frac{11\zeta}{300\sqrt{2}},   \\
\lefteqn{\delta W_{7/2} = -\frac{9}{2}A-\frac{27}{40}B-\frac{27}{5}C -9D}\hspace{11em}\\
 && +\frac{3\eta}{140}+\frac{33\zeta}{140\sqrt{2}}.
\end{eqnarray*}
\begin{eqnarray*}
\lefteqn{C = -\frac{  1}{  30}\delta W_{ 1/2}
      -\frac{  1}{  70}\delta W_{ 3/2} +\frac{  1}{  20}\delta W_{ 5/2}
      }\hspace{11em} \\
      &&-\frac{  5}{ 252}\delta W_{ 7/2} + \frac{\zeta}{350\sqrt{2}} \\
\lefteqn{D =  \frac{  1}{ 210}\delta W_{ 1/2}
      -\frac{  3}{ 490}\delta W_{ 3/2}
   +\frac{  3}{ 980}\delta W_{ 5/2}
     }\hspace{11em}\\
 &&   -\frac{  1}{1764}\delta W_{ 7/2}
\end{eqnarray*}
For $^{25}$Mg, we find that $\eta = 5.37(1)$~MHz and $\zeta=0.333(1)$~MHz,
leading to a value $\Delta C = 0.671(2)$~kHz for the second-order correction to $C$.

$^{43}$Ca, $I=7/2$:
\begin{eqnarray*}
\lefteqn{\delta W_{3/2} =
    -\frac{5}{2}A+\frac{25}{56}B-\frac{55}{7}C +\frac{275}{7}D} \hspace{10em} \\
&&  -\frac{\eta}{112} +\frac{\zeta}{112}\sqrt{\frac{5}{3}},  \\
\lefteqn{\delta W_{5/2} =
 -\frac{7}{2}A+\frac{3}{8}B -44D}\hspace{10em}\\
&&-\frac{\eta}{144} -\frac{\zeta}{48\sqrt{15}}, \\
\lefteqn{\delta W_{7/2} =
    -\frac{9}{2}A+\frac{3}{56}B+\frac{48}{7}C
    +\frac{180}{7}D}\hspace{10em}\\
&&+\frac{\eta}{1680} -\frac{13\zeta}{560}\sqrt{\frac{3}{5}},  \\
\lefteqn{\delta W_{9/2} =
    -\frac{11}{2}A-\frac{33}{56}B-\frac{33}{7}C
    -\frac{55}{7}D}\hspace{10em}\\
&&+\frac{11\eta}{720}+\frac{11\zeta}{240\sqrt{15}}
\end{eqnarray*}
\begin{eqnarray*}
\lefteqn{C = -\frac{  1}{  20}\delta W_{ 3/2 }
      +\frac{  1}{  15}\delta W_{ 7/2 }
 -\frac{  7}{ 220}\delta W_{ 9/2  } } \hspace{16em}\\
 && +\frac{\zeta}{120\sqrt{15}}\\
\lefteqn{D = \frac{  1}{ 140}\delta W_{ 3/2}
      -\frac{  1}{  84}\delta W_{ 5/2 }
   +\frac{  1}{ 140}\delta W_{ 7/2 } }\hspace{16em} \\
  &&    -\frac{  1}{ 660}\delta W_{ 9/2 }
\end{eqnarray*}
For $^{43}$Ca, we find that $\eta = 8.43(3)$~MHz and $\zeta=0.085(1)$~MHz,
leading to a value $\Delta C = -0.183(3)$~kHz for the second-order correction to $C$.

\begin{table*}[ht]
\caption{\label{Tab:others} Theoretical and experimental hyperfine
structure constants for the $^3P_2$ states of $^{9 }$Be, $^{25}$Mg,
and $^{43}$Ca. Theoretical values include the many-body effects
discussed in Section~\ref{Sec:CIMBPT}.}
\begin{ruledtabular}
\begin{tabular}{lldddd}
&& \multicolumn{1}{c}{$A$, MHz} & \multicolumn{1}{c}{$B$, MHz }  &
\multicolumn{1}{c}{$C/\Omega$, $\mathrm{MHz}/(\mu_N \times \text{b})$} &
\multicolumn{1}{c}{$D/\Pi$,  $\mathrm{MHz}/\text{b}^2$ } \\
\hline
$^{9 }$Be & Theory                       & -119.7        & 1.43      & 4.89 \times 10^{-7} &  \\
& Expt.\footnote{Ref.~\cite{BlaLur67}}    & -124.5368(17) & 1.429(8)  & & \\
\\
$^{25}$Mg & Theory                       & -127.5        & 15.8      & 1.71 \times 10^{-5} & -1.39 \times 10^{-14} \\
& Expt.\footnote{Ref.~\cite{Lur62}}       & -128.445(5)   & 16.009(5) & & \\
\\
$^{43}$Ca & Theory                       & -179.9        & -5.50     & 7.03 \times 10^{-5} &  7.83 \times 10^{-13} \\
& Expt.\footnote{Ref.~\cite{GruGusLin79}} & -171.962(2)   & -5.436(8) & & \\
\end{tabular}
\end{ruledtabular}
\end{table*}

\section{Proof that HFS constants $C$ and $D$ may be defined
completely in terms of the HFS intervals} \label{Apdx:Proof} In
this section we prove that the HFS constants $C$ and $D$ can be
expressed uniquely in terms of the HFS intervals even when the
second-order dipole-dipole fine structure term may not be
neglected. This is a nontrivial statement, as the number of linear
equations for HFS intervals is less than the number of fitting
parameters. For example, for $^{87}$Sr we find that there are four
HFS intervals expressed in terms of five fitting parameters (see
Eq.(\ref{Eq:dW-Sr})).

We start by defining our HFS levels from a new energy offset,
$W_F^\prime = W_F + \Delta$. The constant $\Delta$ is arbitrary
and its particular choice will be shown not to affect the
conclusions; consequently, a knowledge of the HFS level intervals
with a convenient choice of $\Delta$ is sufficient to completely
define $W_F^\prime$ for all $F$. Including all first-order terms
as well as the second-order dipole-dipole fine structure term,
the levels $W_F^\prime$ can be written as
\begin{eqnarray}
W_F^\prime &=&
\Delta + \left(-1\right)^{I+J+F}\sum_{k^{\prime}}\left\{
\begin{array}{ccc}
 F & J & I \\
 k^{\prime} & I & J
\end{array}
\right\}Z_{k^{\prime}} \nonumber \\
&& +\left|\left\{ \begin{array}{ccc}  F & J & I \\
 1 & I & J-1
\end{array}
\right\}\right|^2 \eta, \nonumber\\
&&\label{Eq:WFp}
\end{eqnarray}
where
 $Z_k = \langle\gamma J||T_{k}^{e}||\gamma J \rangle\langle I||T_{k}^{n}||I\rangle$.
  From Eq.
(\ref{Eq:HFconsts})
we see that $Z_1$ is proportional to $A$, $Z_2$ is proportional to $B$, etc..
The next step is to multiply every term in Eq. (\ref{Eq:WFp})
by $\left(-1\right)^{I+J+F}\left(2F+1\right)\left\{
\begin{array}{ccc}
 F & J & I \\
 k & I & J
\end{array}
\right\}$, with $k\ne0$, and sum over all $F$-values. Here we analyze the effect of this procedure on the individual terms of the right hand side of Eq. (\ref{Eq:WFp}); to do so, we incorporate various well-known sum rules of six-$j$ symbols. The first term becomes \begin{eqnarray*} \Delta\cdot\sum_{F}\left(-1\right)^{I+J+F}\left(2F+1\right)\left\{
\begin{array}{ccc}
 F & J & I \\
 k & I & J
\end{array}
\right\} \\
=\Delta\cdot\delta_{k,0}\sqrt{\left(2I+1\right)\left(2J+1\right)}=0.
\end{eqnarray*}
The second term becomes
\begin{eqnarray*}
\sum_{k^{\prime}}Z_{k^{\prime}}\sum_{F}\left(2F+1\right)\left\{
\begin{array}{ccc}
 F & J & I \\
 k^{\prime} & I & J
\end{array}
\right\}\left\{\begin{array}{ccc}
 F & J & I \\
 k & I & J
\end{array}
\right\}   \\
= \sum_{k^{\prime}}Z_{k^{\prime}}
\frac{\delta_{k,k^{\prime}}}{\left(2k+1\right)} = \frac{1}{\left(2k+1\right)}Z_{k}.
\end{eqnarray*}
The third term becomes
 \begin{eqnarray*}
\lefteqn{ \eta \left(-1\right)^{2(I+J)+k+1}\sum_{F}\left(-1\right)^{F-I-J-k-1}
\left(2F+1\right) }\hspace{0em} \\
\lefteqn{ \times\left\{
\begin{array}{ccc}
 F & J & I \\
 1 & I & J-1
\end{array}
\right\}\left\{
\begin{array}{ccc}
 F & J & I \\
 1 & I & J-1
\end{array}
\right\}\left\{\begin{array}{ccc}
 F & J & I \\
 k & I & J
\end{array}
\right\} }\hspace{0em} \\
&& =
\eta\left(-1\right)^{2(I+J)+k+1}\left\{
\begin{array}{ccc}
 1 & 1 & k \\
 J & J & J-1
\end{array}
\right\}\left\{
\begin{array}{ccc}
 1 & 1 & k \\
 I & I & I
\end{array}
\right\}.
\end{eqnarray*}
The resulting equation may then be solved for $Z_k$, giving
 \begin{eqnarray*}
\lefteqn{Z_k = \left(2k+1\right)
\sum_{F}\left(-1\right)^{I+J+F}\left(2F+1\right)
\left\{\begin{array}{ccc}
 F & J & I \\
 k & I & J
\end{array}
\right\}W_F^\prime}\hspace{0em} \\
&& +\left(-1\right)^{2(I+J)+k}\left(2k+1\right)  \left\{
\begin{array}{ccc}
 1 & 1 & k \\
 J & J & J-1
\end{array}
\right\}\left\{
\begin{array}{ccc}
 1 & 1 & k \\
 I & I & I
\end{array}
\right\}\eta.
\end{eqnarray*}
First, we note that this expression does not depend on the
specific choice of $\Delta$. Second, we note that for the case of
$k>2$, the triangular condition is not satisfied along the top
rows of the last four six-$j$ symbols in the last expression.
Since these six-$j$ symbols are then equal to zero, $Z_k$ is
completely defined by the values of $W_F^\prime$. Equivalently, we
may conclude that the HFS constants $C, D,...$ may be expressed
completely in terms of the HFS intervals, and these are the same
expressions that would be obtained regardless of the inclusion of
$\eta$.
A more general proof can easily be given to show that with the
inclusion of second-order dipole-quadrupole terms, $C$ can no
longer be expressed completely in terms of the intervals, while
the expression for $D$ in terms of the intervals would still
remain valid (and so on to higher second-order terms if desired).

The above conclusion can also be drawn from the formulation of
second-order effects as in Ref.~\cite{Arm71}, in which the
second-order effects are used to describe the difference between a
measured value of a HFS constant (based on first-order perturbation
theory) compared to its actual value. This already assumes that all
measurable HFI effects are completely described by first- and
second-order perturbation theory. Further assuming that only
second-order terms of the dipole-dipole type contribute to
measurable effects shows that measured HFS constants differ from
actual HFS constants only for the cases of constants $A$ and $B$ and
not for higher constants.

\bibliographystyle{apsrev}

\end{document}